\documentclass[10pt,conference]{IEEEtran}
\IEEEoverridecommandlockouts

\usepackage{cite}
\usepackage{enumerate}
\usepackage{amsmath}
\usepackage[linesnumbered,ruled,vlined]{algorithm2e}
\usepackage{xcolor}
\usepackage[most]{tcolorbox}
\usepackage{booktabs}
\usepackage{threeparttable}
\usepackage{subfigure}
\usepackage{multirow}
\usepackage{url}
\usepackage{balance}
\usepackage{marvosym}

\newcommand{\eg}{\textit{e.g.},\xspace}
\newcommand{\ie}{\textit{i.e.},\xspace}
\newcommand{\etal}{\textit{et al.}}
\newcommand{\wasm}{Wasm\xspace}
\newcommand{\warpdiff}{\textit{WarpDiff}\xspace}
\newcommand{\tool}{\textit{WarpGen}\xspace}
\newcommand{\dis}{\textit{distinguishability}\xspace}
\newcommand{\ds}{\textit{dist score}\xspace}
\newcommand{\pe}{\textit{penalty}\xspace}
\newcommand{\ora}{\textit{oracle ratio}\xspace}
\newcommand{\toolbase}{\textit{WarpGen-base}\xspace}

\def\BibTeX{{\rm B\kern-.05em{\sc i\kern-.025em b}\kern-.08em
    T\kern-.1667em\lower.7ex\hbox{E}\kern-.125emX}}

\begin{document}

\title{Distinguishability-guided Test Program Generation for WebAssembly Runtime Performance Testing}

\author{
\IEEEauthorblockN{Shuyao Jiang\textsuperscript{*}, Ruiying Zeng\textsuperscript{\dag}, Yangfan Zhou\textsuperscript{\dag,\Letter}\thanks{\textsuperscript{\Letter} Yangfan Zhou is the corresponding author.} and Michael R. Lyu\textsuperscript{*}}
\IEEEauthorblockA{\textsuperscript{*} \textit{Department of Computer Science and Engineering, The Chinese University of Hong Kong, Hong Kong, China}}
\IEEEauthorblockA{\textsuperscript{\dag} \textit{School of Computer Science, Fudan University, Shanghai, China}}
\IEEEauthorblockA{syjiang21@cse.cuhk.edu.hk, ryzeng22@m.fudan.edu.cn, zyf@fudan.edu.cn, lyu@cse.cuhk.edu.hk}
}

\maketitle

\begin{abstract}
  WebAssembly (Wasm) is a binary instruction format designed as a portable compilation target, which has been widely used on both the web and server sides in recent years. 
  As high performance is a critical design goal of Wasm, it is essential to conduct performance testing for Wasm runtimes.
  However, existing research on Wasm runtime performance testing still suffers from insufficient high-quality test programs.
  To solve this problem, we propose a novel test program generation approach \textit{WarpGen}.
  It first extracts code snippets from historical issue-triggering test programs as initial operators, then inserts an operator into a seed program to synthesize a new test program.
  To verify the quality of generated programs, we propose an indicator called \textit{distinguishability}, which refers to the ability of a test program to distinguish abnormal performance of specific Wasm runtimes. 
  We apply \textit{WarpGen} for performance testing on four Wasm runtimes and verify its effectiveness compared with baseline approaches. In particular, \textit{WarpGen} has identified seven new performance issues in three Wasm runtimes.
\end{abstract}

\begin{IEEEkeywords}
WebAssembly, performance testing, test program generation
\end{IEEEkeywords}

\section{Introduction}\label{sec_intro}

WebAssembly (\wasm)~\cite{haas2017bringing} is a binary instruction format designed as a portable compilation target for programming languages (\eg C, C++ and Rust).
\wasm is designed to execute at native speed, provide type and memory safety, and be portable across different languages and platforms.
These advantages of \wasm make it popular in a wide range of fields, including web and non-web environments~\cite{lehmann2023sa,lehmann2022finding,lehmann2020everything,lehmann2019wasabi,romano2023automated}.
Four major browsers (\ie Chrome, Firefox, Safari, and Edge) have supported \wasm to enable higher-performance web applications~\cite{wagner2017webassembly}.
\wasm has also been increasingly used in many server-side applications (\eg cloud service~\cite{shillaker2020faasm,gackstatter2022pushing,gadepalli2020sledge}, microcontrollers~\cite{gurdeep2019warduino,zandberg2021femto}, and smart contracts~\cite{zheng2020vm,wang2020wana,chen2022wasai}).

\textit{High performance} is a critical design goal of \wasm, so it is highly essential to conduct \wasm runtime performance testing, especially for server-side \wasm applications.
Unlike web applications running on well-developed browser engines, server-side \wasm applications need to run in standalone \wasm runtimes (\eg WasmEdge~\cite{wasmedge}), which are still immature and more likely to cause performance issues (\ie abnormal latency).
Jiang \etal~\cite{jiang2023revealing} recently pointed out the severe impact of performance issues on server-side \wasm applications. They found that a 30ms-latency will result in up to 50\% drop of service throughput in a real-world \wasm microservice~\cite{microservice}.
They further proposed a differential testing approach \warpdiff to identify some performance issues in existing standalone \wasm runtimes. The results indicate that performance issues are common in many \wasm runtimes, which can significantly threaten the reliability of \wasm applications.

However, state-of-the-art research on \wasm runtime performance testing still suffers from insufficient high-quality test programs.
\warpdiff only used a small benchmark (\ie 123 programs from the LLVM Test Suite~\cite{llvmtest}) for testing, so the identified issues are very limited.
To further uncover more performance issues, it is necessary to collect or generate more test programs.
Unfortunately, existing test suites and test program generation approaches are mainly targeted at finding functional bugs (\ie software errors that cause wrong execution results) in specific software (\eg Csmith~\cite{yang2011finding}, a popular program generator for C compiler testing). 
They are unsuitable for applying directly to \wasm runtime performance testing since the test programs are not tailored for triggering \wasm runtime performance issues.
Without high-quality (\ie issue-triggering) test programs, much time and manual effort will be spent on case filtering and issue verification, dramatically hurting testing efficiency.




To solve this problem, it is critical to design an \textit{efficient} test program generation approach for \wasm runtime performance testing.
However, it is extremely challenging to achieve this goal. Specifically, we need to address two main challenges.
The first challenge is the lack of prior knowledge about what programs tend to trigger performance issues in \wasm runtimes.
Unlike other testing tasks (\eg compiler testing and JVM testing) that have been widely studied for decades, \wasm runtime performance testing is still in the early stages of practice. 
To the best of our knowledge, the only current experience with \wasm runtime performance is the few cases reported by \warpdiff, which is too limited for test program generation.
The second challenge is how to verify the quality of newly generated test programs.
We treat test programs that trigger performance issues in \wasm runtimes as high-quality test programs. However, it is infeasible to verify whether each generated program triggers performance issues manually. We need to design an indicator to automatically verify the quality of test programs and keep high-quality ones.

In this paper, we propose \textbf{\tool}, a novel test program generation approach for \wasm runtime performance testing.
To address the first challenge, \tool adopts the idea of history-driven test program generation. 
The insight is that historical issue-triggering test programs contain code snippets that help detect new issues. 
This idea has been verified in many testing tasks, including C compiler testing~\cite{chen2019history,rabin2021configuring} and JVM testing~\cite{zhao2022history}.
Based on this idea, our limited prior knowledge about \wasm runtime performance can still provide insights for finding new issues. 
Specifically, \tool first extracts code snippets from abnormal cases (\ie the test programs that have revealed \wasm runtime performance issues) reported by \warpdiff. Then, \tool integrates these code snippets with different contexts to generate new test programs for triggering new performance issues.
However, not all generated programs can achieve this goal. 
Therefore, to address the second challenge, we propose an indicator, namely \textbf{\dis}, to measure the quality of the generated test programs.
Conceptually, the \dis refers to the ability of a test program to distinguish the abnormal performance of some \wasm runtimes.
To formalize it, we draw inspiration from the idea of \warpdiff, that is, the execution time of the same test case on different \wasm runtimes should follow a stable ratio (\ie \ora). So, we can identify an abnormal case whose execution time ratio significantly deviates from the \ora.
Hence, we formalize the \dis of a test program as the distance between the vector of its execution time ratio on several \wasm runtimes and the vector of the \ora, called \ds.
\tool uses the \ds to measure the quality of a test program and guide the whole process of program generation.

Based on the above insights, the workflow of \tool is designed as follows: 
For preparation, \tool extracts a series of code snippets (called \textit{operators}) from the abnormal cases reported by \warpdiff to construct the operator pool, and collects some seed programs to construct the seed pool.
In the initial iteration, for each operator, \tool inserts it into a seed program to generate a new test program, then executes the \wasm code on several \wasm runtimes and calculates its \ds.
After the initial iteration, \tool treats the test programs with top $N$ \ds as \textit{distinguishable programs}, then extracts operators from these programs and adds them to the operator pool. 
To improve the efficiency of further iterations, \tool assigns a \pe with an initial zero value for each operator.
In the follow-up iterations, each time, \tool randomly selects an operator and a seed program to generate a new program, then calculates its \ds after execution.
If the \ds exceeds the previous top $N$, \tool will mark it as a new \textit{distinguishable program}. Also, \tool will update the top $N$ scores and reset the \pe of the selected operator.
Otherwise, the \pe of the selected operator will increase by one. 
When the \pe of an operator accumulates to $M$, it will be removed from the operator pool. 
Thus, \tool can generate programs with an increasing trend of \ds.

For evaluation, we conducted a series of experiments.
We used \tool to generate C programs for performance testing on four popular \wasm runtimes: Wasmer~\cite{wasmer}, Wasmtime~\cite{wasmtime}, WasmEdge~\cite{wasmedge}, and WAMR~\cite{wamr}.
To evaluate the efficiency of \tool to generate high-quality test programs, we collected the statistics of \ds during the program generation.
We found that the \ds of \textit{distinguishable programs} achieved the optimal result quickly, which means that \tool can generate high-quality test programs efficiently.
To further verify the effectiveness of the \dis-guided design, we compared \tool with a random approach Csmith and a non-guided approach \toolbase.
We compared the \ds of the generated test programs from the three approaches, and we found that \tool can achieve the best overall performance.
Finally, \tool identified seven previously unknown performance issues in three \wasm runtimes. 
The results verify the effectiveness of \tool for \wasm runtime performance testing.

In summary, this work makes the following contributions:
\begin{itemize}
    \item \textbf{Direction.} We conduct the first study on test program generation for \wasm runtime performance testing, and we propose a novel indicator (\ie \dis) to guide test program generation.
    \item \textbf{Approach.} We design and implement a \dis-guided test program generation approach \tool, which aims to efficiently generate test programs that trigger performance issues in \wasm runtimes.
    \item \textbf{Empirical study.} We apply \tool for performance testing on four real-world \wasm runtimes and verify its effectiveness compared with other approaches. \tool has identified seven previously unknown performance issues in three \wasm runtimes.
\end{itemize}

We make our source code and experiment data available at \url{https://figshare.com/s/9d1b5e43b80029d14c42}.

\section{Background}\label{sec_bg}

\subsection{\wasm Runtime}

\begin{figure}[t]
    \centering
    \includegraphics[width=0.95\linewidth]{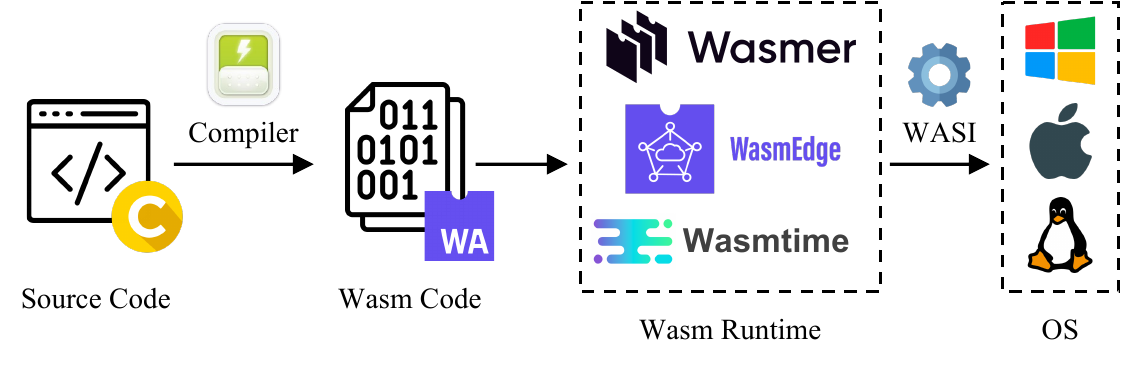}
    \vspace{-0.5em}
    \caption{The workflow of \wasm compilation and execution.}
    \label{fig_wasm}
\end{figure}

\wasm is designed as a compilation target for high-level programming languages (\eg C/C++). The typical workflow of deploying a \wasm application is first to compile the source program into \wasm bytecodes, then execute the program on a \wasm runtime, as shown in Figure~\ref{fig_wasm}.
The \wasm runtime plays a vital role in this workflow.
It is a virtual machine that translates \wasm binary instructions to native machine code for execution. 
It provides a memory-safe, sandboxed execution environment for \wasm, working as an intermediary between the \wasm application and the operating system (OS).
The mechanism of \wasm runtime enables \wasm to be portable across different platforms.

For \wasm applications in different environments, the implementation of \wasm runtimes is different.
In the web environment, the \wasm bytecodes run in the browser with the help of JavaScript glue code. The glue code would call into the browser engine (\eg V8~\cite{v8}, SpiderMonkey~\cite{spidermokey}), which would then talk to the operating system. 
However, in the non-web (server-side) environment, \wasm needs a particular interface, the WebAssembly system interface (WASI)~\cite{clark2019standardizing}, to access the resources in the operating system. So, server-side \wasm applications need to run in a standalone \wasm runtime with WASI support.
With the increasing applications of \wasm in the non-web environment, many standalone \wasm runtimes are emerging. There are currently more than 30 standalone \wasm runtime implementations on GitHub~\cite{awesome}.
Some representative standalone \wasm runtimes include Wasmer~\cite{wasmer}, Wasmtime~\cite{wasmtime}, WasmEdge~\cite{wasmedge}, WAMR~\cite{wamr}, etc.

Unlike the major browser engines that are well-developed, existing standalone \wasm runtimes are still in the early stages of development.
Most standalone \wasm runtimes do not have mature optimization mechanisms, which may cause performance issues (\ie abnormal latency) during execution.
Compared to functional issues, performance issues in \wasm runtimes are more likely to be overlooked but can lead to severe consequences.
Jiang \etal~\cite{jiang2023revealing} have found that a short latency of 30ms will result in up to 50\% drop of service throughput in a real-world \wasm microservice~\cite{microservice}.
Thus, performance testing is a critical task for \wasm runtimes.

\subsection{Performance Testing for \wasm Runtimes}

Although performance testing is critical for \wasm runtimes, research in this area is still limited.
Existing works about \wasm performance mainly study the systematic performance gaps between \wasm and native code or JavaScript~\cite{jangda2019not,wang2021empowering,yan2021understanding,spies2021evaluation}, but little focus on \textit{performance issues} in \wasm runtimes.
One state-of-the-art research towards \wasm runtime performance issues is \warpdiff~\cite{jiang2023revealing}, a differential testing approach for identifying performance issues in \wasm runtimes.

\warpdiff aims to solve the problem of lacking test oracle in performance testing on \wasm runtimes, \ie how to identify the occurrence of performance issues.
Unlike functional issues with a clear test oracle (\eg wrong execution results), performance issues are hard to detect since there is no ground truth of the performance indicator (\ie the execution time of a test case).
To solve this problem, \warpdiff proposed a new idea for performance testing: \textit{The execution time of the same test case on different \wasm runtimes should follow a stable ratio (\ie \textbf{\ora}) under normal circumstances.}
For each test case, \warpdiff collected its execution time on several \wasm runtimes to construct a vector representing its execution time ratio on these runtimes. Then, it got the vector of the \ora based on the average execution time ratio of all the test cases.
\warpdiff compared the distance between the two normalized vectors for each test case, then identified an abnormal case in which the execution time ratio significantly deviates from the \ora.
Based on this idea, \warpdiff conducted performance testing on several \wasm runtimes using 123 test programs from the LLVM Test Suite~\cite{llvmtest} and identified seven performance issues in four runtimes.

Although \warpdiff has addressed the lack of test oracle for \wasm runtime performance testing, it still suffers from another challenge: insufficient high-quality test programs.
\warpdiff only used a small benchmark for testing. Thus, the number of identified performance issues is also limited.
To discover more performance issues in \wasm runtimes, it is necessary to conduct large-scale performance testing using more test programs.
However, using randomly selected programs for performance testing is inefficient, since most test cases cannot trigger performance issues in \wasm runtimes.
Without high-quality test programs, we must spend a lot of time and manual effort on issue verification.
Therefore, to improve the testing efficiency, we aim to design a test program generation approach that produces high-quality (\ie issue-triggering) test programs for \wasm runtime performance testing.

\section{Approach}\label{sec_app}


\begin{figure*}[t]
    \centering
    \includegraphics[width=0.95\linewidth]{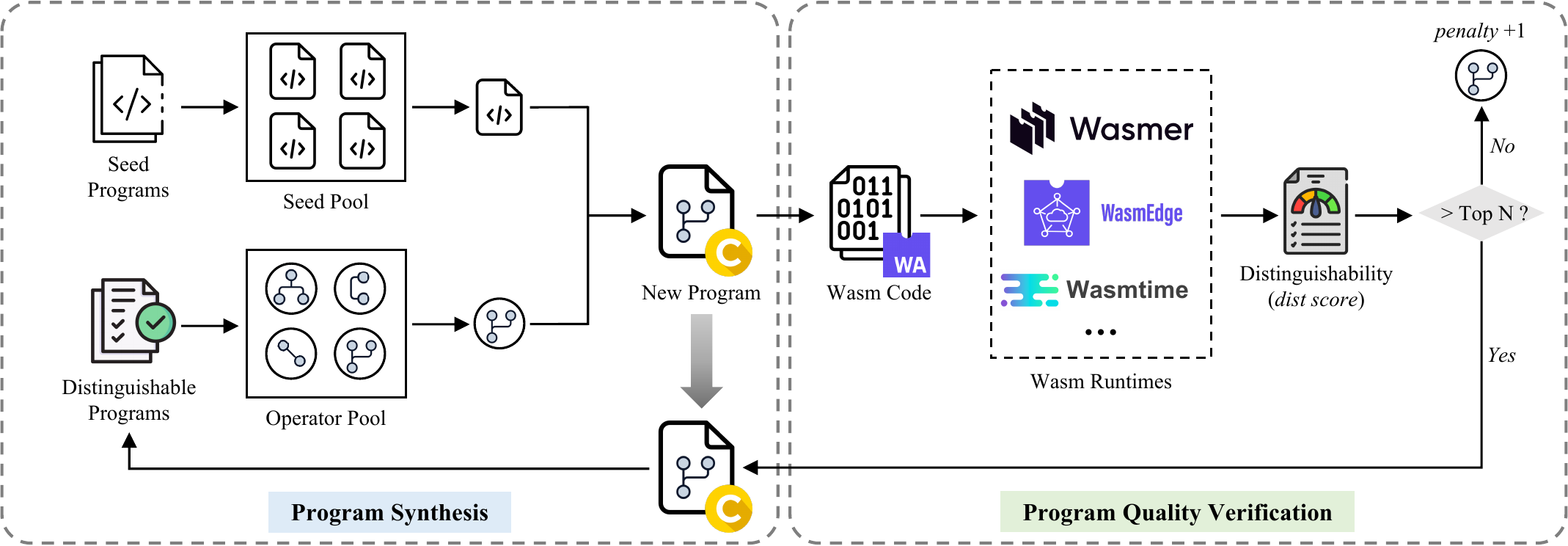}
    \caption{The framework of \tool.}
    \label{fig_overview}
\end{figure*}

Although test program generation has been widely studied, there is still a lack of test program generation approach targeting at \wasm runtime performance testing.
There are two unique challenges in this task:
1) \textit{We lack sufficient prior knowledge about what kinds of programs can trigger performance issues.} Compared with other testing tasks (\eg compiler testing) that have been studied for many years, our experience on \wasm runtime performance is too limited for test program generation.
2) \textit{It is difficult to verify the quality of generated test programs.} The key criterion for high-quality test programs is that they can trigger performance issues in \wasm runtimes, \ie they can distinguish the abnormal performance of some \wasm runtimes. However, it is infeasible to check each test program manually. We need an indicator for automatic quality verification of test programs.

To address the above challenges, we propose \textbf{\tool}, a novel test program generation approach for \wasm runtime performance testing.
The design of \tool contains two key insights:
1) \textit{Historical issue-triggering test programs contain information that helps detect new issues. }
This idea has been verified in many previous studies, including compiler testing~\cite{chen2019history,rabin2021configuring} and JVM testing~\cite{zhao2022history}.
Although our historical experience towards \wasm runtime performance issues is limited (only a few known issues reported by \warpdiff), we can still extract useful information from those previous issue-triggering test programs to generate new test programs.
2) \textit{The test oracle proposed by \warpdiff can inspire test program quality verification.}
\warpdiff has verified that \wasm runtime performance issues can be revealed from those test programs in which the execution time ratio significantly deviates from the \ora. 
In other words, such test programs can distinguish the abnormal performance of certain \wasm runtimes and should be considered high-quality test programs.
Therefore, we propose an indicator called \textbf{\dis} to measure the quality of the generated programs. 
We formalize the \dis of a test program as the distance between the two vectors of its execution time ratio and the \ora, called \ds.
\tool uses the \ds to guide the test program generation process.

Figure~\ref{fig_overview} shows the framework of \tool.
Overall, \tool contains two key modules: program synthesis and program quality verification.
As a preparation, \tool first extracts a series of code snippets (called \textit{operators}) from the historical issue-triggering test programs reported by \warpdiff to initialize the operator pool; then collects some seed programs to initialize the seed pool (\textbf{Section~\ref{subsec_pre}}).
To generate a new test program, \tool selects an operator to insert into a seed program. During the insertion, we need to solve two critical problems: \textit{insertion point selection} and \textit{variable dependency handling} (\textbf{Section~\ref{subsec_syn}}).
The iteration process is designed as follows:
In the initial iteration, \tool first generates a new test program for each initial operator, then collects the test programs with top $N$ \ds as \textit{distinguishable programs} for updating the operator pool.
In the follow-up iterations, for a newly generated test program, \tool checks whether its \ds exceeds the previous top $N$ for further updates (\textbf{Section~\ref{subsec_ite}}). 
Next, we will elaborate on the design and implementation details of \tool.

\subsection{Data Preprocessing}\label{subsec_pre}

In this subsection, we first introduce how \tool preprocesses data for test program generation, including \textit{operator extraction} and \textit{seed profiling}.
Since the operators and seeds are from different programs, a big challenge of synthesizing a new program is to ensure the validity of the synthesized program. The program validity includes two aspects: 
\begin{itemize}
    \item \textbf{Syntax Validity:} The synthesized program should conform to the syntax rules and be able to pass compilation.
    \item \textbf{Insertion Validity:} The inserted operator should affect the behavior of the seed program.
\end{itemize}
To address this challenge, the key operation in data preprocessing is to record the related contexts for program synthesis, which will be explained later.

\textbf{Operator Extraction.}
To extract operators (code snippets) from a source program, we first need to determine the operator granularity, \eg line, block, or function.
Fine-grained code snippets (\eg code line) have simple context but usually cannot express complete functionality. On the contrary, coarse-grained code snippets (\eg function) may have a complex context that is difficult to process during program synthesis.
Therefore, \tool adopts block granularity for extraction since it can achieve a trade-off between functional completeness and processing difficulty.
Specifically, \tool extracts the following four types of operators:
\begin{itemize}
    \item \textbf{Sequential Operator:} A code block containing sequential statements without branches and loops.
    \item \textbf{Branching Operator:} A code block containing conditional branching statements (\ie \texttt{if}, \texttt{else}, and \texttt{else if}), including conditions and corresponding bodies.
    \item \textbf{Looping Operator:} A code block containing looping statements (\ie \texttt{for}, \texttt{while}, and \texttt{do-while}), including loop conditions and loop bodies. 
    \item \textbf{Mixed Operator:} A code block containing a combination of the above three operators.
\end{itemize}

For the implementation of operator extraction, \tool implements a customized Clang tool named \texttt{extract-blocks} based on the LLVM Project~\cite{llvm-project}, a modular toolchain for C-like languages.
For each input source program, \tool first transfers the program to an Abstract Syntax Tree (AST).
The AST describes the syntactical structure of the source program. Each node in the AST refers to a statement (\texttt{Stmt}) or a declaration (\texttt{Decl}) in the program, and each edge refers to the logic between the two nodes.
To extract operators from the source program, \tool traverses its AST to collect \texttt{Stmt} with specific classes.
Specifically, to extract branching operators, \tool collects statements with the class of \texttt{IfStmt}. To extract looping operators, \tool collects statements with the classes of \texttt{ForStmt}, \texttt{WhileStmt}, and \texttt{DoStmt}.
For sequential operators, we cannot collect them directly since there is no class representing sequential statements in AST. To solve this problem, \tool first collects statements with the class of \texttt{CompoundStmt}, which refers to a block surrounded by `\texttt{\{\}}'. Among these blocks, those without branching statements and looping statements are treated as sequential operators, and the rest are treated as mixed operators.

As mentioned above, a critical challenge in program synthesis is to ensure the \textit{syntax validity} and \textit{insertion validity} of the new program.
To achieve this goal, during operator extraction, \tool records the following two contexts for each operator:
\begin{itemize}
    \item \textbf{Pre-context:} All used variables (excluding variables declared and assigned in this block) and called functions (excluding standard library functions) in this block. 
    \item \textbf{Post-context:} All variables assigned in this block (excluding variables declared in this block). 
\end{itemize}

On the one hand, the variables and functions in the pre-context would not be found in the seed program and should be replaced accordingly to ensure the \textit{syntax validity}.
We exclude the declared variables and the standard library functions because they would not cause syntax errors in the synthesized program.
On the other hand, the variables in the post-context should be replaced by the variables that would be used after the insertion point of the operator. Only in this way can the inserted operator affect the behavior of the seed program to ensure the \textit{insertion validity}.
To obtain the two contexts of an operator, \tool first traverses all its used variables and functions in the AST (nodes of type \texttt{DeclRefExpr}). For each variable, \tool checks whether it is on the left side of an \texttt{AssignmentOperator} (\eg \texttt{=}, \texttt{+=}, and \texttt{-=}). If so, the variable will be added to the post-context. Otherwise, it will be added to the pre-context. The functions (excluding standard library functions, \eg \texttt{printf}) will be directly added to the pre-context.
For each variable and function, \tool records its name, type, and all positions.

\begin{figure}[t]
    \centering
    \includegraphics[width=0.98\linewidth]{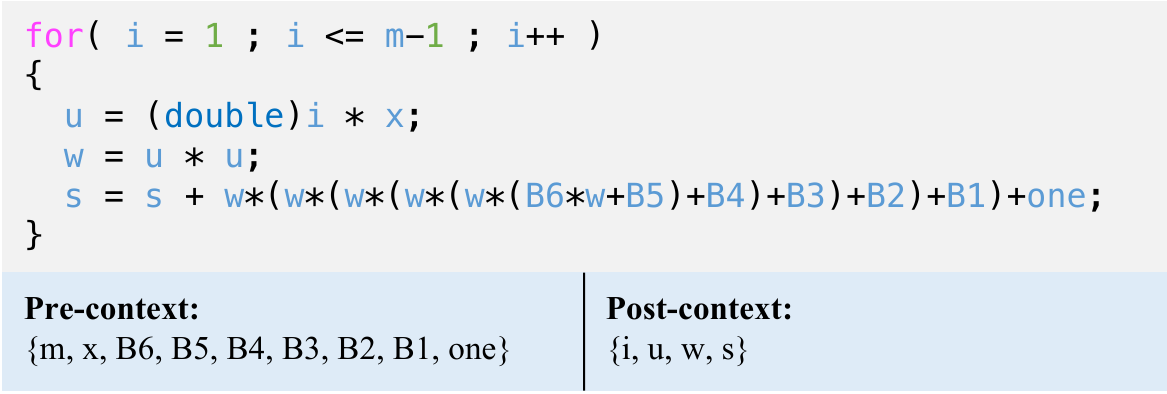}
    \caption{A \textit{looping operator} and its pre/post-contexts. The type and positions of the variables are omitted.}
    \label{fig_context}
\end{figure}

To better explain how \tool obtains the two contexts, we provide an example, as shown in Figure~\ref{fig_context}.
This example shows a looping operator extracted by \tool.
In the source code of this operator, four variables (\ie \texttt{i}, \texttt{u}, \texttt{w}, and \texttt{s}) are assigned new values. To ensure the \textit{insertion validity}, the values of these four variables should be used after the insertion point of this operator.
Therefore, \tool adds them to the post-context of this operator.
Other variables used but not declared in this code (\ie \texttt{m}, \texttt{x}, \texttt{B6}, \texttt{B5}, \texttt{B4}, \texttt{B3}, \texttt{B2}, \texttt{B1}, and \texttt{one}) are added into the pre-context. They should be replaced or redefined to ensure the \textit{syntax validity}.

After initial operator extraction, \tool will obtain some initial operators from the historical issue-triggering test programs. 
Each operator is saved as a JSON file, recording its source code and the pre/post-contexts.
During the iteration process of \tool, it will further extract new operators from new \textit{distinguishable programs} to update the operator pool.

\textbf{Seed Profiling.}
To provide diverse contexts for synthesized programs, \tool picks some C programs different from the source programs of the operators to construct the seed pool. The selection of seed programs can be customized.
To ensure the \textit{syntax validity} and \textit{insertion validity} of the synthesized programs, \tool also needs to profile the seed programs.

Specifically, \tool profiles the \textit{variable usage} and \textit{code coverage} for each seed program. 
The \textit{variable usage} is used to ensure both \textit{syntax validity} and \textit{insertion validity}. 
Given a seed program, \tool first traverses its AST to collect all the declared variables and functions (nodes of type \texttt{VarDecl}). For each variable (including function), \tool records the positions (code line number) where the variable was first defined and last used. 
The purpose of recording the two positions is: 
During program synthesis, the variables for replacing the pre-context of the operator should be defined before the insertion point, and the variables for replacing the post-context of the operator should be used after the insertion point.
\tool implements a Clang tool named \texttt{ues-define-tag} to obtain the \textit{variable usage}. 
On the other hand, the \textit{code coverage} is used to ensure the \textit{insertion validity}. To make a valid insertion, the insertion point of the operator should always be covered during program execution, \ie the inserted operator can always be executed.
Therefore, given a seed program, \tool also executes it and profiles its \textit{code coverage} using \texttt{llvm-profdata} and \texttt{llvm-cov}~\cite{llvm-project}, then records all the covered code lines during execution.

\subsection{Program Synthesis}\label{subsec_syn}
In this subsection, we describe how \tool synthesizes a new test program given an operator and a seed program.
During program synthesis, \tool needs to resolve two critical problems: \textit{insertion point selection} and \textit{variable dependency}.
\tool resolves these two problems based on the preprocessing information described in Section~\ref{subsec_pre}. The specific process is as follows:

\textbf{Insertion Point Selection.}
For the same seed program, the valid insertion points. (\ie the synthesized program can satisfy the \textit{syntax validity} and \textit{insertion validity}) of different operators are different, since each operator has unique pre/post-contexts.
Therefore, before inserting an operator, \tool first needs to search valid insertion points for this operator.
To this end, \tool traverses the code lines recorded in the \textit{code coverage} information of the seed program (described in Section~\ref{subsec_pre}) and checks which lines can be valid insertion points.
Specifically, when \tool visits a specific code line, it collects the current context of the seed program, including all the local variables that have been defined at this point and all the global variables. This current context is obtained based on the \textit{variable usage} of the seed program.
Then, \tool reads the pre/post-contexts of the operator to check whether the current context of the seed program can meet the requirements for replacing the pre/post-contexts of the operator.
For each variable (including function) to be replaced, \tool first checks if there is a reusable local/global variable in the seed program. If not, \tool will check whether it is feasible to define a new variable here.
In this way, \tool finds all valid insertion points for the operator and randomly selects a point for insertion.

\textbf{Variable Dependency Handling.}
After selecting an insertion point, \tool then handles the variable dependency of the inserted operator.
As mentioned above, for a variable (including function) $x$ to be replaced, \tool first tries to find a reusable variable $y$ in the seed program. 
The reusable variable $y$ should satisfy the following conditions: 
First, The type of $y$ must be the same as $x$. 
Meanwhile, for a variable in the post-context of the operator, the last used position of $y$ should be after the current insertion point.
If such a variable $y$ is found, \tool will replace $x$ at all positions in the operator with $y$.
If multiple variables satisfy the condition, \tool will randomly select one for replacement.

However, if no reusable variables for $x$ are found in the seed program, \tool will define a new variable to ensure the validity of the synthesized program.
Specifically, \tool first declares a new variable $n$ of the same type as $x$, then assigns a value for $n$. The assigned value is randomly selected from a predefined variable pool.
\tool has implemented the function of defining new variables for all basic data types, array and pointer types in C.
After defining the variable $n$, \tool will replace $x$ at all positions in the operator with $n$, then insert the variable definition statements and the modified operator into the seed program to synthesize the new program.
\tool implements a Clang tool named \texttt{insert} to achieve the above program synthesis process.

\subsection{Iteration Process}\label{subsec_ite}
In this subsection, we elaborate on the whole iteration process of \tool.
In particular, we will explain how \tool adopts \dis (\ds) as the program quality indicator to guide the test program generation process.

Based on our design insights, those test programs that can distinguish the abnormal performance of certain \wasm runtimes are treated as high-quality test programs.
Such test programs are manifested as that, their execution time ratio on several \wasm runtimes significantly deviates from the \ora.
Therefore, we propose an indicator named \dis to measure the quality of test programs. 
We formalize the \dis of a test program as \ds, which is defined as follows:


\begin{tcolorbox}[colback=gray!10,
                  colframe=black,
                  width=\linewidth,
                  arc=3mm, auto outer arc,
                  boxrule=1pt
                 ]
  \small{\textit{The \textbf{\dis (\ds)} of a test program is the Euclidean distance between the normalized vector of its execution time ratio (on \wasm runtimes to be tested) and the normalized vector of the \ora.}}
\end{tcolorbox}

For example, if the execution time of a test program $x$ on three \wasm runtimes are 1s, 2s, and 3s, then its original execution time vector can be represented as $[1,2,3]$.
However, the original vectors of different test programs are not comparable since the execution time of each program is affected by its features.
Therefore, we need to normalize the original execution time vector for each program. In this way, the test programs with the same execution time ratio will have the same normalized vectors. For example, the original vector $[10,20,30]$ of test program $y$ will be normalized to $[0.17,0.33,0.5]$ (\textit{L1-normalization}), the same as that of $x$.
If the normalized vector of the \ora is $[0.2,0.4,0.4]$, then the \ds of $x$ (and also $y$) is 0.12, the Euclidean distance between the two normalized vectors.
The higher \ds means the test program is more capable of distinguishing abnormal performance in specific \wasm runtimes.

\begin{algorithm}[t]
\SetKwFunction{SynthesizeProgram}{synthesizeProgram}
\SetKwFunction{GetDistScore}{getDistScore} 
\SetKwFunction{ExtractOps}{extractOps} 
\SetKwFunction{Min}{min} 
\SetKwFunction{Update}{update} 
\SetKwInOut{Input}{Input}
\SetKwInOut{Output}{Output}
\SetKwRepeat{Do}{do}{while}

\footnotesize{
	\Input{$seedPool$, $opPool$, $N$, $M$, $k$} 
	\Output{top $N$ \textit{distinguishable programs}}
	 \BlankLine 

      \tcp{Initial Iteration} 
      \ForEach{operator $op$ in $opPool$}{
        $seed \leftarrow$ a random seed in $seedPool$\;
        $synProgram \leftarrow$ \SynthesizeProgram{$op$, $seed$}\;
        $distScore \leftarrow$ \GetDistScore{$synProgram$}\;
      } 
      $topScoreSet \leftarrow$ top $N$ $distScore$ values\;
      $topProgramSet \leftarrow$ programs with top $N$ $distScore$ values\;
      \ForEach{program $p$ in $topProgramSet$}{
        $opPool \leftarrow$ $opPool$ $\cup$ \ExtractOps{$p$}\;
      }
                  
      \lForEach{operator $op$ in $opPool$}{
        $penalty_{op} \leftarrow 0$    
      }
      \tcp{Follow-up Iterations} 
      \While{program numbers $ < k$ \texttt{AND} $opPool \neq \emptyset$}{
        $op, seed \leftarrow$ a random value in $opPool, seedPool$\;
        $synProgram \leftarrow$ \SynthesizeProgram{$op$, $seed$}\;
        $distScore \leftarrow$ \GetDistScore{$synProgram$}\;
        \If{$distScore >$ \Min{$topScoreSet$}}{
            $opPool \leftarrow$ $opPool$ $\cup$ \ExtractOps{$synProgram$}\;
            \Update{$topScoreSet$, $topProgramSet$}\;
            $penalty_{op} \leftarrow 0$\;
        }
        \lElse{$penalty_{op} \leftarrow penalty_{op}+1$}
        \lIf{$penalty_{op} == M$}{$opPool \leftarrow opPool \setminus \left\{op\right\}$}
      }
}   
 
\caption{Iteration Process of \tool}
 \label{algo_warp}
\end{algorithm}
\DecMargin{1em}

Next, we will describe the whole iteration process of \tool, as shown in Algorithm~\ref{algo_warp}.
The iteration process aims to identify the top $N$ \textit{distinguishable programs}, \ie test programs with top $N$ \ds (the value of $N$ can be customized). 
The intention is that the performance of the tested \wasm runtimes deviates more from the expected performance (\ie \ora) when executing these programs, \ie they can reveal performance issues with high severity.

\textbf{Initial Iteration.}
In the initial iteration, \tool aims to set the initial top $N$ \textit{distinguishable programs}.
To this end, \tool first traverses the initial operator pool. For each operator, \tool tries to insert it into a randomly selected seed program to synthesize a new test program (based on the strategy described in Section~\ref{subsec_syn}). 
If an operator cannot be inserted into any seed program successfully, \tool will remove it from the operator pool.
Then, \tool compiles the synthesized program to \wasm code using Emscripten~\cite{zakai2011emscripten}, a typical compiler for compiling C/C++ programs to \wasm.
After compilation, \tool executes the \wasm code on several \wasm runtimes (test objects) and calculates the \ds of this test program.
Thus, after the initial iteration, \tool collects the test programs with top $N$ \ds as the initial \textit{distinguishable programs}.
However, we cannot identify which part of these test programs caused their high \ds without manual analysis.
A reasonable solution is to extract operators from these test programs for further iterations.
Therefore, at this step, \tool extracts operators from the initial \textit{distinguishable programs} and adds them to the operator pool. 

\textbf{Follow-up Iterations.}
In follow-up iterations, the goal of \tool is to continuously find new test programs with higher \ds than the previous top $N$ programs.
According to our design, the operators in the operator pool will gradually increase during the iteration process, but not every operator is effective for generating high-quality test programs. Random selection of operators is inefficient for generating high-quality test programs.
Therefore, to improve the efficiency of test program generation, we design a \textit{penalty mechanism} for operator selection.
Before iterations, \tool assigns a \pe with an initial zero value for each operator (including those obtained in the initial iteration) in the operator pool.
During the follow-up iterations, each time, \tool randomly selects an operator and a seed to generate a new test program, then calculates its \ds after execution. 
If the \ds exceeds the previous top $N$ scores, \tool will mark it as a new \textit{distinguishable program}, then extract its operators and add them to the operator pool. 
Also, \tool will update the top $N$ scores and reset the \pe of the selected operator to zero.
Otherwise, the \pe of the selected operator will be increased by one. 
When the \pe of an operator is accumulated to $M$, it will be removed from the operator pool. The value of $M$ can also be customized.
After multiple iterations, the operators remaining in the operator pool will be more likely to contribute to test programs with high \ds.
Finally, the iteration process will stop when the total number of generated test programs reaches the customized value $k$, or when the operator pool becomes empty.

\section{Evaluation}\label{sec_eval}




In this section, we address the following research questions:
\begin{itemize}
    \item \textbf{RQ1:} How efficient is \tool to generate high-quality test programs?
    \item \textbf{RQ2:} How effective is the \dis-guided design in \tool?
    \item \textbf{RQ3:} Can \tool detect new performance issues in existing \wasm runtimes?
\end{itemize}

\subsection{Experiment Settings}

\textbf{Test objects.}
In our experiments, we selected four standalone \wasm runtimes, \ie Wasmer~\cite{wasmer}, Wasmtime~\cite{wasmtime}, WasmEdge~\cite{wasmedge}, and WAMR~\cite{wamr}, as test objects.
Table~\ref{tab_runtimes} shows their information. 
The considerations for selecting these runtimes include two aspects.
On the one hand, they are all representative \wasm runtimes with high \textit{popularity} (\ie with the top number of GitHub stars) and \textit{activity} (\ie the latest commit is within one month).
On the other hand, \warpdiff has reported that all these four \wasm runtimes contain performance issues, which need our further attention. 
We selected the latest version of each runtime for testing and tested all the runtimes in AOT (Ahead-of-Time) mode.

\textbf{Initial Operators.}
Based on the first insight of \tool, We collected historical issue-triggering test programs to initialize the operator pool. 
Specifically, we collected the top 20 abnormal test programs reported by \warpdiff for operator extraction, since they are currently the only source related to \wasm runtime performance issues.
These programs are from four Benchmarks (\ie BenchmarkGame, Misc, Shootout, and Polybench) in the LLVM Test Suite~\cite{llvmtest} and written in C.
Finally, \tool obtained 271 initial operators from these programs. 
We show the details of these operators in our supplementary data due to space limitations.


\begin{table}[t!]
\centering
\caption{\wasm runtimes as test objects.}
\vspace{-0.5em}
\begin{threeparttable}
\begin{tabular}{ccccc}
 \toprule
 \textbf{Runtime} & \textbf{Language} & \textbf{\#Stars}  & \textbf{\#Commits} & \textbf{Version} \\
 \midrule
 Wasmer & Rust & 16.7k & 16.3k & 4.2.3 \\
 Wasmtime & Rust & 13.5k & 12.5k & cli 15.0.0 \\
 WasmEdge & C/C++ & 7.2k & 2.9k & 0.13.5 \\
 WAMR & C/C++ & 4.2k & 1.5k & 1.2.3 \\
 \bottomrule
\end{tabular}
\end{threeparttable}
\label{tab_runtimes}
\vspace{-1em}
\end{table}

\begin{table*}[t!]
\centering
\caption{Statistics of top 20 \ds when \tool generated different numbers of test programs.}
\begin{tabular}{lccccccccccccccc}
 \toprule
 \textbf{\#Programs} & 20  & 50 & 80 & 110 & 140 & 170 & 200 & 230 & 260 & 290 & 320 & 350 & 380 & 410 & 436\\
 \midrule
 Minimal & 0.023 & 0.149 & 0.208 & 0.259 & 0.454 & 0.454 & 0.454 & 0.454 & 0.454 & 0.454 & 0.454 & 0.458 & 0.459 & 0.459 & 0.459 \\
 Average & 0.145 & 0.226 & 0.266 & 0.392 & 0.462 & 0.462 & 0.462 & 0.462 & 0.462 & 0.462 & 0.462 & 0.479 & 0.481 & 0.481 & 0.481\\
 \bottomrule
\end{tabular}
\label{tab_dist}
\end{table*}

\textbf{Seed Programs.}
Since the initial operators are code snippets from C programs, the seeds should also be C programs.
In our experiments, we selected 100 C programs randomly generated by Csmith~\cite{yang2011finding} as seeds.
Csmith is one of the most popular random program generators that can generate C programs free of undefined behaviors, which is suitable for providing seed programs in our task.
It is worth noting that \tool accepts any C program as a seed. So, there are also many other ways of setting seed programs, and we just chose one appropriate method for experiments.

\textbf{Parameters.}
There are some customized parameters in \tool design.
First, \tool needs to maintain the top $N$ values of \ds and the corresponding test programs during iteration. We set the value of $N$ to 20 in experiments, because the performance deviation shown by the cases after the top 20 is less significant according to the data reported by \warpdiff.
Second, we set the \pe threshold $M$ for operator removal to 5, which can achieve a trade-off between iteration efficiency and test program quality based on our observation during the experiments.
Then, we set the total number $k$ of generated test programs to 1,000 in our experiments.
In addition, we calculated the \ora based on the average execution time ratio of the seed programs, which represents the normal performance of the tested \wasm runtimes.

\textbf{Compared Approaches.}
To answer RQ2, we compared \tool with two baseline approaches. 
The first one is random programs (different from the seeds) generated by Csmith.
The second one is a simplified version of \tool (named \toolbase), where the \dis-guided design is removed. \toolbase generates a test program just by randomly selecting an operator and a seed for program synthesis. It will not update the operator pool and the top program set.
We evaluated the quality of the test programs under these three approaches to verify the effectiveness of the \dis-guided design in \tool.

\textbf{Implementation and Environment.}
We implemented \tool as a Java project, which integrates all of our self-implemented Clang tools (\ie \texttt{extract-blocks}, \texttt{use-define-tag}, and \texttt{insert}, written in C++) and other necessary tools for testing (\ie \wasm compiler and runtimes). The parameters and the data paths can be easily modified in the configuration file. 
We also implemented a complete error handling and checkpoint mechanism for \tool.
The project is implemented by over 5k lines of code, including 2k lines of Java and 3k lines of C++.
All our experiments are conducted on a server with a 4-core Intel(R) Xeon(R) E5-2686 v4 CPU @ 2.30GHz and 16GB RAM. The operating system is 64-bit Ubuntu 20.04.6 LTS with the Linux kernel version 5.15.0.

\subsection{RQ1: Efficiency of \tool}
To evaluate the efficiency of \tool to generate high-quality test programs, we collected the statistics of \ds for each generated test program during the iteration process of \tool.
For analysis, we counted the minimal of the top 20 \ds (\ie the 20th \ds) and the average of the top 20 \ds with the number of generated test programs increased.
We reported these data because the goal of \tool is to find \textit{distinguishable programs} with top $N$ \ds. Since we set the value of $N$ to 20 in our experiments, we focused on the top 20 \ds, which indicates the quality of the target test programs of \tool.
We reported the minimal of the top 20 \ds since it is the threshold for updating \textit{distinguishable programs}. The change in this value reflects the growth trend of the test program quality.
We also reported the average of the top 20 \ds, which reflects the average quality of the \textit{distinguishable programs}.

The results are shown in Table~\ref{tab_dist}.
We report the results for every 30 test programs generated by \tool.
We finally got 436 generated programs in our experiments, since the operator pool became empty before the number of generated test programs reached the set value (early stop).
It is worth mentioning that the early stop does not mean the effectiveness of \tool is not good. On the contrary, it means that \tool can achieve the optimal result quickly without generating a large amount of test programs.
We can observe from Table~\ref{tab_dist} that, in the early stage of test program generation (\ie when the number of generated programs is less than 150), the minimal and the average of top 20 \ds grew quickly.
When \tool generated 140 test programs, the minimal of the top 20 \ds had reached 0.454, approaching the final optimal result of 0.459.
The results indicate that \tool can generate high-quality test programs for \wasm runtime performance testing with high efficiency.

\begin{figure}[t]
  \centering
  \subfigure[Minimal of top 20]{\includegraphics[width=0.49\linewidth]{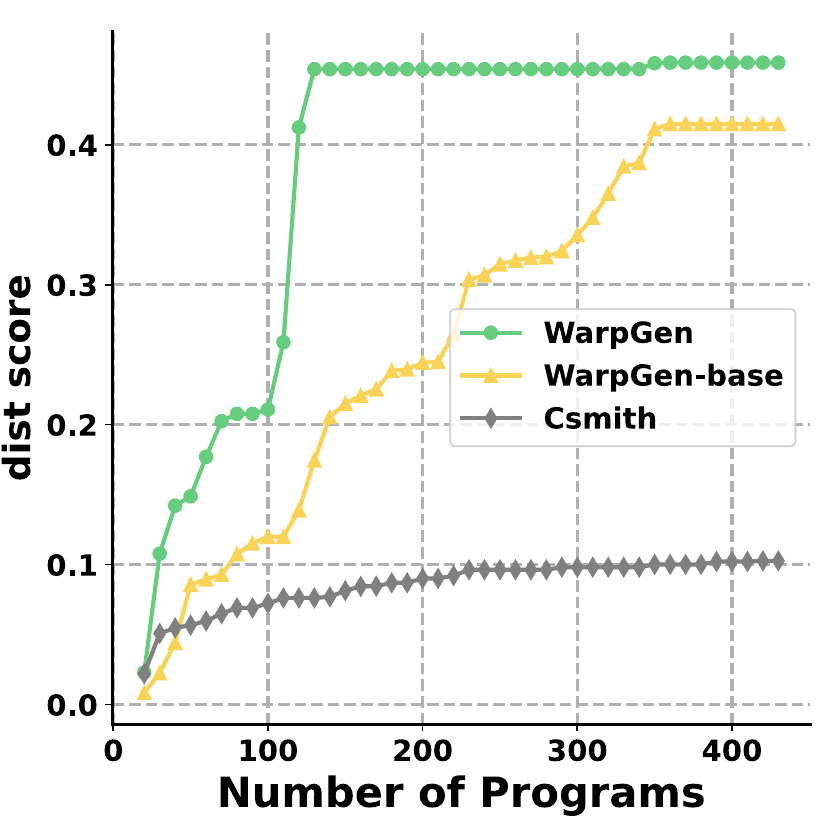}}
  \subfigure[Average of top 20]{\includegraphics[width=0.49\linewidth]{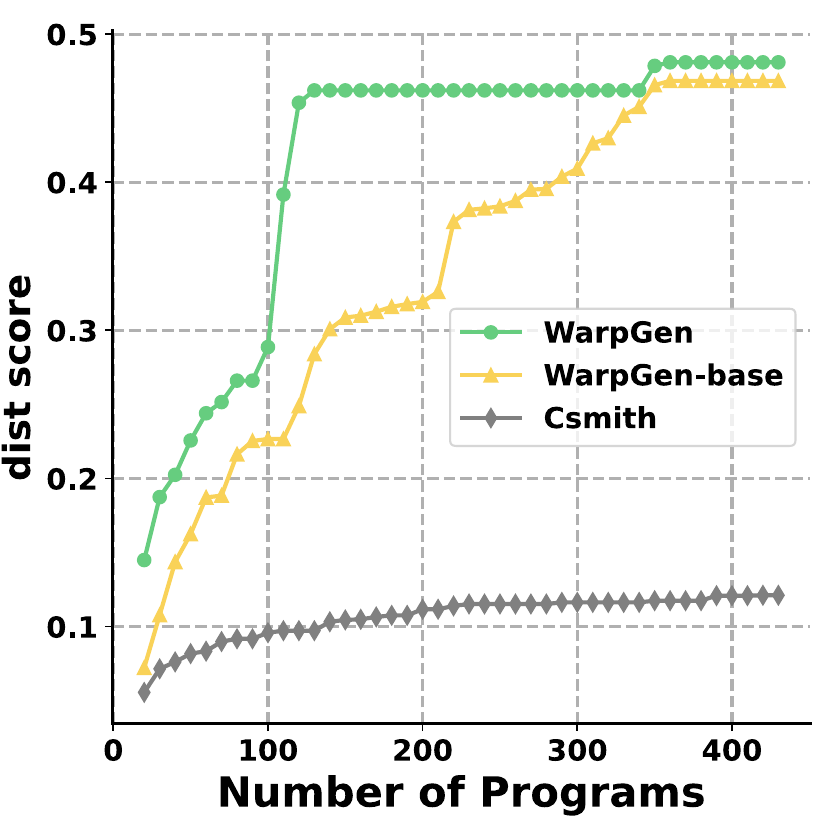}}
  \caption{Comparison of top 20 \ds from different approaches.}
  \label{fig_dist}
\end{figure}

\subsection{RQ2: Effectiveness of Guidance}
To further study the effectiveness of \tool, especially the \dis-guided design, we compared \tool with two baseline approaches, Csmith~\cite{yang2011finding} and \toolbase.
Csmith is a representative random C program generator. We treated Csimth as a baseline to evaluate the effectiveness of \tool relative to random approaches.
\toolbase is a simplified version of \tool without guidance in the program generation process. We implemented \toolbase to evaluate the effectiveness of the \dis-guided design in \tool.
We applied these two baseline approaches to generate the same number of test programs as \tool, and compared the change in the top 20 \ds from the three approaches. Similar to RQ1, we reported the minimal and the average of the top 20 \ds. The comparison results are shown in Figure~\ref{fig_dist}, where the data was sampled for every ten programs generated.

From the comparison results, we can find that \tool achieved the best performance among the three approaches.
On the one hand, the \textit{growth rate} of \tool's curve is significantly greater than that of the two baseline approaches in the early stage of program generation. \tool achieved the near-optimal result quickly (when generating 100+ programs), while \toolbase did not achieve the optimal result until generating 350+ programs. For Csmith, the top 20 \ds changed little during the whole program generation process. This result indicates that \tool can generate high-quality test programs \textit{more quickly} than the baseline approaches.
On the other hand, the \textit{final results} of \tool are also the best among the three approaches, for both the minimal and the average of the top 20 \ds. In particular, the final results of \tool are four times greater than that of Csmith. This result indicates that \tool can also generate \textit{higher-quality} test programs than the baseline approaches.

In summary, \tool can generate test programs of \textit{higher quality} and \textit{more quickly} than the baseline approaches.
The comparison between \tool and Csmith shows the effectiveness of \tool relative to random approaches.
The comparison between \tool and \toolbase verifies that the \dis-guided design is effective for \tool to generate high-quality test programs with high efficiency.

\subsection{RQ3: New Performance Issues}

\begin{table}[t!]
\centering
\caption{Performance issues identified by \tool.}
\begin{threeparttable}
\resizebox{\linewidth}{!}{
\begin{tabular}{cccc}
 \toprule
 \textbf{ID} & \textbf{Runtime}  & \textbf{Scenario} & \textbf{Status}\\
 \midrule
 \#7731 & Wasmtime & Floating-point (FP) arithmetic & Fixed \\
 \#7732 & Wasmtime & Access of pointers to constant & Confirmed \\
 \#7733 & Wasmtime & Increment operation in nested loops & Confirmed \\ 
 \midrule
 \#4378 & Wasmer & Operations on FP arrays & Fixed \\
 \#4379 & Wasmer & Call of standard output functions & Fixed \\
 \#4380 & Wasmer & Access of variable addresses & Fixed \\
 \midrule
 \#2938 & WAMR & FP arithmetic & Confirmed \\
 \bottomrule
\end{tabular}
}
\end{threeparttable}
\label{tab_issues}
\end{table}

To verify the ability of \tool to identify new performance issues in \wasm runtimes, we analyzed the top 20 \textit{distinguishable programs} generated by \tool.
We first need to locate the issue-related runtime for each test program, \ie determine in which \wasm runtime the abnormal latency occurred.
To this end, we adopted the performance issue location method in \warpdiff. 
Given the execution results of a \textit{distinguishable program}, we analyzed the impact of each \wasm runtime on this anomaly respectively, based on the execution time of this program on each runtime.
Specifically, for each dimension in the normalized execution time vector, we calculated its distance from the value of the same dimension in the normalized vector of the \ora. This distance is called \textit{deviation degree}, representing the impact of the corresponding \wasm runtime on this anomaly. Thus, we treated the \wasm runtime with the largest \textit{deviation degree} as the issue-related runtime.
Then, we analyzed the scenarios where these performance issues occurred based on the source code of the related test programs, and we verified the issues by reproducing them on similar test programs.
Finally, we summarized seven performance issues in three \wasm runtimes, which are all previously unknown issues. The results are shown in Table~\ref{tab_issues}.
Currently, all the issues are confirmed by developers, and four issues have been fixed.

For Wasmtime, we identified three performance issues.
Issue \#7731 occurred when Wasmtime handling floating-point (FP) arithmetic. 
The inserted operator in the issue-related test program is from a historical abnormal test program, which contains multiple FP arithmetic expressions.
It is worth noting that the historical test program did not reveal this issue in the previous study. This verifies the effectiveness of the first design insight of \tool, \ie the historical issue-triggering test programs can help detect new issues.
Issue \#7732 is related to the improper optimization when accessing the pointers to constant. In this case, the inserted operator is from a newly generated test program, where the anomaly occurred when accessing a variable of type `\texttt{const int32\_t *}'. This result also verifies the effectiveness of the \dis-guided design of \tool, \ie the newly extracted operators also help detect performance issues.
Issue \#7733 was caused by an increment operation (\ie \texttt{x++}) in nested loops. 

For Wasmer, we also identified three performance issues.
Issue \#4378 occurred when Wasmer executed the operations on FP arrays, which was reflected in multiple test programs. Specifically, we found that abnormal latency occurred on the operations of arrays with types of \texttt{float[]}, \texttt{double[]}, and \texttt{double*[]}.
Issue \#4379 reflects the improper handling of Wasmer when calling the standard output function \texttt{fprintf()}.
Issue \#4380 was caused by insufficient optimization when accessing variable addresses. 

For WAMR, we identified one performance issue (Issue \#2938) related to insufficient FP arithmetic optimization, similar to Issue \#7731 in Wasmtime. This issue was also reflected in multiple test programs, one with the same inserted operator as Issue \#7731. This result indicates that the same inserted operator in different contexts (\ie seed programs) can trigger performance issues in different \wasm runtimes.

In summary, the results verify the ability of \tool to detect new performance issues in \wasm runtimes.
The results also show the effectiveness of the key design insights (\ie history-driven and \dis-guided) in \tool, which indicates that \tool is a practical test program generation approach for \wasm runtime performance testing.

\section{Discussion}\label{sec_dis}

\subsection{Case Study}

\begin{figure}[t]
    \centering
    \includegraphics[width=\linewidth]{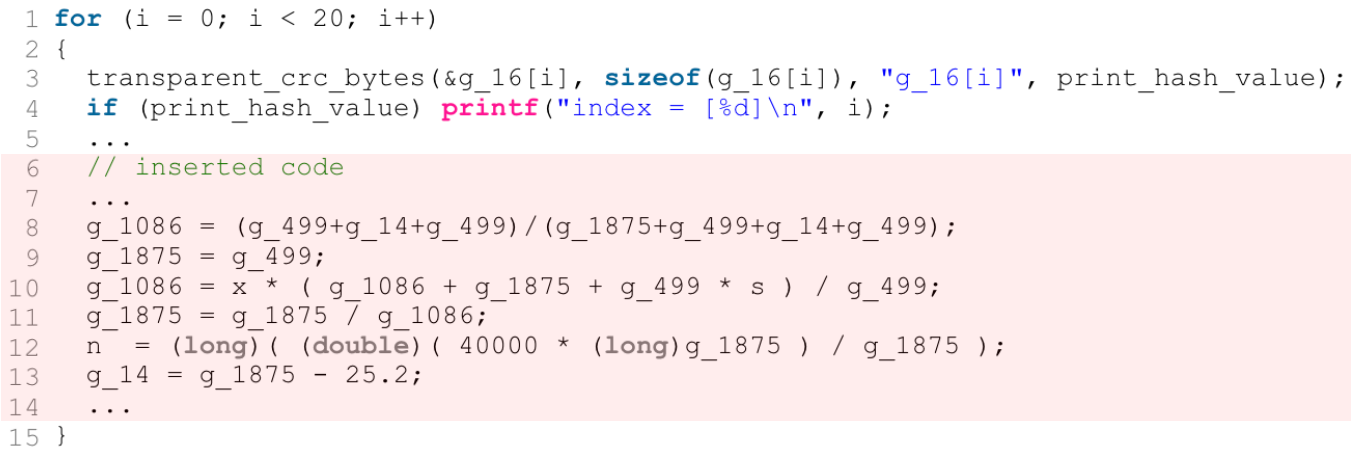}
    \caption{Test program related to Issue \#7731.}
    \label{fig_exp}
\end{figure}

In this subsection, we provide a case study to show the effect of \tool intuitively. Due to space limitations, we just show one representative case and leave others in issue reports.

Figure~\ref{fig_exp} shows part of the code in the test program related to Issue \#7731.
The inserted operator in this test program was extracted from a historical abnormal case \texttt{flops-1.c} reported by \warpdiff. It contains several FP arithmetic expressions, where the original variables were all replaced with same-type variables in the seed program (\ie variables whose names begin with `\texttt{g\_}', type of \texttt{double}).
This operator was inserted in a loop (\texttt{for} statement) of the seed program.
When this test program was executed, Wasmtime experienced severe abnormal latency, which was about three times slower than its expected performance.
However, this issue has not been discovered in the previous study. We found that this issue cannot be triggered when Wasmtime executes \texttt{flops-1.c} alone. It only occurred in the specific context of this generated test program. 
We further investigated the reason and found that the inserted code was executed in a loop, amplifying the abnormal performance. Meanwhile, the modified variables changed the original logic of the seed program, which also contributed to this anomaly.

This case study indicates that performance issues in \wasm runtimes may be triggered under various unexpected conditions, which cannot be detected by existing test programs.
\tool can produce diverse issue-triggering test programs to detect more performance issues in \wasm times with high efficiency.
Therefore, \tool is a promising tool for \wasm runtime performance testing.

\subsection{Threats to Validity}


The threats to \textit{internal} validity mainly lie in the implementation of \tool.
To reduce the threats, we carefully checked all the code and conducted sufficient testing. For the Clang tools of operator extraction and program synthesis, we used large-scale C programs to verify their correctness. For the Java framework of \tool, we conducted multiple unit and integration tests for each module. We also implemented a comprehensive log system to monitor its workflow.

The threats to \textit{external} validity mainly lie in the programs used for operator extraction and as the seeds.
\tool extracted the initial operators from the abnormal test programs reported by \warpdiff since they are the only source of historical issue-triggering test programs we know.
For the seed programs, we selected some random C programs generated by Csmith~\cite{yang2011finding} in our experiments. Csmith is one of the most widely used C program generators, so it can provide a representative result. 

The threats to \textit{construct} validity mainly lie in the settings of parameters in the experiments.
First, the parameter $N$ affects the threshold for updating the \textit{distinguishable programs} in \tool. We set $N$ to 20 in our experiments based on the experience of \warpdiff, since the anomalies reflected in subsequent test programs are less significant.
Second, the parameter $M$ affects the speed of removing an operator. Based on the observations in multiple pre-experiments, we set $M$ to 5 since it can achieve a trade-off between iteration efficiency and test program quality. 

\section{Related Work}\label{sec_re}

\textbf{\wasm Runtime Performance.}
\wasm~\cite{haas2017bringing} is a binary instruction format serving as a compilation target for programming languages~\cite{lehmann2023sa,lehmann2022finding,lehmann2020everything,lehmann2019wasabi,romano2023automated}. 
\wasm's fast, safe, lightweight, and portable features make it popular on both web side~\cite{reiser2017accelerate,hilbig2021empirical,romano2022wobfuscator} and server side~\cite{mendki2020evaluating,makitalo2021webassembly,kjorveziroski2022evaluating,nurul2021nomad}.
High performance is a critical design goal of \wasm. Regarding \wasm runtime performance, existing studies mainly focus on the web side~\cite{jangda2019not,wang2021empowering,yan2021understanding,de2022webassembly,de2021runtime,romano2023function}.
For example, Jangda \etal~\cite{jangda2019not} designed BROWSIX-\wasm to conduct the first large-scale evaluation of the performance of \wasm vs. native. 
Wang~\cite{wang2021empowering} investigated how browser engines optimize \wasm execution in comparison to JavaScript.
Yan \etal~\cite{yan2021understanding} further extended this study to three major browser engines (Chrome, Firefox, and Edge).
Romano and Wang~\cite{romano2023function} recently investigated the counterintuitive impacts of function inlining on \wasm runtime performance on the web.

There are also a few studies on server-side \wasm runtime performance.
Spies and Mock~\cite{spies2021evaluation} evaluated \wasm runtime performance in non-web environments. Their measurements demonstrated that \wasm is generally faster than JavaScript.
Recently, Jiang \etal~\cite{jiang2023revealing} identified the significant impact of performance issues on server-side \wasm runtimes, and they proposed a differential testing approach \warpdiff to reveal performance issues in server-side \wasm runtimes.
Some other studies~\cite{wang2023comprehensive,zhang2023characterizing, liu2023exploring,romano2021empirical} investigated \wasm runtime bugs and showed several cases related to runtime performance.
However, state-of-the-art research on \wasm runtime performance testing still suffers from insufficient high-quality test programs, which we aim to solve in this work.

\textbf{Test Program Generation.}
High-quality test programs are critical for software testing. Test program generation aims to construct effective and diverse test programs, typically applied to compiler testing~\cite{boujarwah1997compiler,chen2017learning,chen2019compiler,chen2019history,chen2020survey, wu2023jitfuzz,tu2022remgen} and JVM testing~\cite{chen2016coverage,chen2019deep,brennan2020jvm,zhao2022history,gao2023vectorizing}.
Existing program generation approaches can be generally divided into two categories: generation-based approaches and mutation-based approaches.
Csmith~\cite{yang2011finding} and YARPGen~\cite{livinskii2020random} are two typical program generators designed for generating C programs free of undefined behaviors.
Alipour \etal~\cite{alipour2016generating} proposed directed swarm testing~\cite{groce2012swarm} that uses statistics and a variation of random testing to produce random tests.
HiCOND~\cite{chen2019history} and K-Config~\cite{rabin2021configuring} extract insights from historical bug reports and use the insights to guide the test program generators.
Typical mutation-based approaches include \textit{equivalence modulo inputs} (EMI)~\cite{le2014compiler}, Athena~\cite{le2015finding}, and Hermes~\cite{sun2016finding}, which produce program variants by mutating code snippets in existing programs.
CLsmith~\cite{lidbury2015many} follows the idea of EMI to validate OpenCL compilers by mutating dead code in test programs.
Donaldson and Lascu~\cite{donaldson2016metamorphic} further proposed strategies for metamorphic testing of OpenGL compilers using opaque value injection.
There are also some machine learning-based program generation approaches, \textit{e.g.}, DeepSmith~\cite{cummins2018compiler} and DeepFuzz~\cite{liu2019deepfuzz}.

Regarding \wasm runtime performance testing, a significant challenge is the lack of sufficient test programs.
Although some studies aimed to generate \wasm code~\cite{hassler2021wafl,lehmann2021fuzzm,zhou2023wadiff,zhao2024wapplique}, none are targeted at runtime performance testing.
So, in this work, we propose the first test program generation approach for \wasm runtime performance testing, improving the efficiency of discovering more performance issues in \wasm runtimes.

\section{Conclusion}\label{sec_con}
In this paper, we propose \tool, a novel test program generation approach for \wasm runtime performance testing.
It first extracts code snippets from historical issue-triggering test programs as operators, then inserts an operator into a random seed to generate a new test program. It adopts an indicator \dis to verify the quality of test programs and guide the program generation process.
Our experiments have shown that \tool can generate high-quality test programs more efficiently than other baseline approaches. \tool has identified seven previously unknown performance issues in three \wasm runtimes.
The results verify the effectiveness of \tool for \wasm runtime performance testing.


\section*{Acknowledgment}
This work was supported by the Natural Science Foundation of Shanghai (No. 22ZR1407900) and the RGC Grant for Theme-based Research Scheme Project (RGC Ref. No. T43-513/23-N).

\balance
\bibliographystyle{IEEEtran}
\bibliography{reference}

\end{document}